\documentclass[12pt,preprint]{aastex}
\usepackage{CJK}
\shortauthors{Sadjadi et al.} 
\shorttitle{Alphatic side groups}

\received{}
\usepackage{graphicx}
\usepackage{epsfig}

\graphicspath{{converted_graphics/}}
\begin{document}

\title{ON THE ORIGIN OF THE 11.3 MICRON UNIDENTIFIED INFRARED EMISSION FEATURE
}
\begin{CJK*}{Bg5}{bsmi}
\CJKtilde

\author{SeyedAbdolreza Sadjadi,  Yong Zhang (±iªa),  and Sun Kwok (³¢·s)}
\affil{Space Astronomy Laboratory, Faculty of Science, The University of Hong Kong, Pokfulam Road, Hong Kong, China}
\email{sunkwok@hku.hk} 

\begin{abstract}

The 11.3 $\mu$m emission feature is a prominent member of the family of unidentified infrared emission (UIE) bands and is frequently attributed to out-of-plane bending modes of polycyclic aromatic hydrocarbon (PAH) molecules.  We have performed quantum mechanical calculations of 60 neutral PAH molecules and found that it is difficult to reconcile the observed astronomical feature with any or a mix of these PAH molecules.  We have further analyzed the fitting of  spectra of several astronomical objects by the NASA PAH database program and found that reasonable fittings to the observed spectra are only possible by including significant contributions from oxygen and/or magnesium containing molecules in the mix.  A mixed of pure PAH molecules, even including units of different sizes, geometry and charged states, is unable to fit the astronomical spectra. Preliminary theoretical results on the vibrational spectra of simple molecules with mixed aromatic/aliphatic structures show that these structures have consistent bundles of vibrational modes and could be viable carriers of the UIE bands.


\end{abstract}


\footnote {Accepted for publication in the Astrophysical Journal May 14, 2015}

\clearpage

\section{Introduction}

The unidentified infrared emission (UIE) bands at 3.3, 6.2, 7.7, 8.6, and 11.3, and 12.7\,$\mu$m are a family of infrared emission bands which are widely observed in planetary nebulae, reflection nebulae, H{\sc ii} regions, diffuse interstellar clouds, and galaxies.  In star forming
galaxies, UIE bands contribute to up to 20\% of the total infrared output of the galaxies
\citep{sd07}.  The carrier of the UIE features therefore must be a major constituent of interstellar matter and their  spectral characteristics can serve as a potential probes for the chemical processes and physical conditions of interstellar environments \citep[][]{jt11,pe13}. 


The 11.3\,$\mu$m feature is a prominent member of the UIE family and was among the first UIE bands detected \citep{gf73}.  The detection of this feature in the planetary nebulae NGC 7027, BD+30\degr3639, and NGC 6572 came as a surprise as only atomic fine-structure lines were expected to be present in these objects.  
The strength of the 11.3 $\mu$m feature was found not to be related to the excitation conditions
of the nebulae and its origin was first attributed to mineral carbonates such as 
MgCO$_3$ \citep{gf73}.  This identification was soon discarded when the expected 
7.0\,$\mu$m feature from carbonates was not found \citep{rs77b}.  
\citet{rs77a} found the 11.3\,$\mu$m feature to be closely correlated with the 3.3\,$\mu$m feature, suggesting a common origin for the two features. The current accepted
explanation is that these  two features arise from the C--H vibration modes of aromatic compounds  \citep{dw81}.  Many aromatic compounds are known by laboratory spectroscopy to show absorption bands around 3 $\mu$m due to C--H stretch and around 12 $\mu$m due to C--H out-of-plane (OOP) bending modes.
The fact that the  11.3\,$\mu$m feature is found almost exclusively in  planetary nebulae with high C to O ratio ($>2$), and sometimes associated with the faint SiC feature, suggests carbon grains as the carrier \citep{ba83}.  Since then, various carbonaceous  materials containing an aromatic component have been suggested as the UIE carrier.  The list of candidates includes hydrogenated amorphous carbon \citep[HAC,][]{duley1983}, quenched carbon composites \citep[QCC,][]{sak87}, polycyclic aromatic hydrocarbon (PAH) molecules \citep{lp84,al85,pl89}, coal \citep{papoular1989}, petroleum fractions \citep{cataldo2003, cataldo2013a}, and mixed aromatic/aliphatic organic nanoparticles \citep[MAONs,][]{kz11, kz13}. 
Among these proposals, the PAH hypothesis has captured the most attention and is widely believed as the carrier of the UIE features by the astronomical community.

Assuming a PAH origin, the 11.3\,$\mu$m feature  has been frequently used as a tool to probe
the physical conditions of galaxies and AGNs \citep[e.g.,][]{sd07,gt08,wh10}.
The 11.3\,$\mu$m feature is used to trace the neutral and/or large PAHs, and the strength ratios between 
the 11.3\,$\mu$m feature and shorter wavelength (such as 6.2, 7.7, and 8.6\,$\mu$m) UIE bands have been used to estimate the properties of the radiation fields. 
The physical validity of these studies, however, depends on a correct interpretation of the origin of the 11.3 $\mu$m feature.  

Recent advances in quantum-chemical methodology and computational power provide the accuracy needed to quantitatively study the UIE phenomenon. The NASA Ames PAH IR Spectroscopic Database \citep[PAHdb;][]{bb10,boersma2014} collects theoretical infrared spectra of 700
PAH molecules.  Using the density functional theory (DFT), we have calculated the IR spectra of a series of molecules with mixed aliphatic-aromatic structures \citep{sadjadi2015}. In this study, we reexamine the 11.3\,$\mu$m feature utilizing the PAHdb data and new quantum chemistry calculations.

\section{Observations of the 11.3 $\mu$m feature in different astronomical sources}


The observed peak position of the 11.3\,$\mu$m feature in astronomical sources is well defined  and  does not vary much in wavelength.  
The 11.3 $\mu$m  feature  is also detected in absorption \citep{bh00},  and its wavelengths and profiles closely resemble those seen in emission. The 11.3 $\mu$m feature has a distinctive asymmetric profile, having a steep decline in the short wavelength side and a gradual extended wing in the long wavelength side \citep{vp04}.  Such asymmetric profiles are difficult to explain by gas-phase molecular emissions.  Explanations offered include having PAH molecules of different mass ranges, with the high-mass PAHs responsible for the short-wavelength side and low-mass  PAHs for the long-wavelength side \citep{candian2015}.  
Furthermore, the peak positions of the C--H OOP bending modes of PAH molecules can be quite different due to mode coupling, molecular structure and charge states of the molecules \citep{hv01}.  Because of the large wavelength variations of the OOP modes of PAH molecules, it is difficult to assign or match the observed astronomical feature to specific PAH molecules. One suggestion of a possible solution to  this problem is to incorporate  internal hydrogen in  carbonaceous microparticles  \citep{bk90}. Ring deformation vibrational mode of small carbonaceous molecules such as ethylene oxide ($c$-C$_2$H$_4$O) has also been proposed to explain the narrow 11.3\,$\mu$m feature \citep{bl09}.

Like the other UIE bands, the 11.3 $\mu$m feature is not seen in AGB stars but only emerges during the post-AGB phase of evolution.
 \citet{gc95} did not detect this feature in 718 carbon stars, except a feature at 
 11.9\,$\mu$m  in several sources. The situation is further complicated
by the blending of a SiC feature \citep[``class $\delta$'' spectra according to][]{mb14}.
A sample study of Herbig Ae/Be stars shows that the central wavelength of the 11.3\,$\mu$m feature decreases with increasing effective temperature of the central stars \citep{ks08}, suggesting that the chemical structure of its carrier is being  processed by stellar 
radiation.  In the spectral classification scheme of \citet{peeters02},
the 11.3\,$\mu$m feature occurs at  longer wavelength in Class C sources in comparison to those in class A and B objects.
Since Class C sources usually show aliphatic feature (e.g., at 3.4 $\mu$m), it is thought that the peak position of the 11.3\,$\mu$m feature may reflect the aliphatic/aromatic ratio.

The 11.3 $\mu$m feature is not an isolated spectral feature.  In the spectra of proto-planetary nebulae, weaker emission features at 12.1, 12.4 and 13.3 $\mu$m are also seen to accompany the 11.3 $\mu$m feature  \citep{kwok1999}. 
It is known from laboratory studies of substituted aromatic molecules that the frequencies of the OOP bending mode correlates with the number of adjacent H atom on each ring \citep{bellamy1958}. This empirical rule has been translated to the correlation between OOP vibrational frequencies and the number of exposed hydrogens in PAH molecules \citep{hudgins1999}.  In the simplest PAH molecule benzene, all corners are exposed and the molecule has 6 exposed H atoms.  As the number of fused rings increases in two dimensions in a plane (growth of graphene flakes), the number of exposed H on each benzene ring varies.  In general, the more compact a PAH molecule, there are fewer number of exposed H sites on each benzene ring of a PAH molecule.   The hydrogens on these benzene units are thus labeled as solo, duo, trio, quartet, quintet and sextet corresponding to 1, 2, 3, 4 ,5 or 6 exposed H corners.  It has  been suggested that the wavelength of OOP mode increases monotonically from solo to sextet \citep{hudgins1999}.  
The observed emission features at 11.3, 12.1, 12.4 and 13.3 $\mu$m may therefore  correspond to the OOP modes of solo, duo, trio, and quartet systems.  We should note that these OOP bending modes are also seen in more complex amorphous hydrocarbons with mixed aromatic/aliphatic structures \citep{herlin1998}.   Absorption features at 11.36, 11.92, 12.27, and 13.27 $\mu$m in the laboratory sample are identified as solo, duo, trio, and quartet OOP modes respectively.

In addition to the narrow features mentioned above, astronomical spectra in this spectral region also show a broad emission plateau extending from  11 to 13\,$\mu$m.  This plateau feature has been suggested to be due to OOP bending modes of a collection of PAH molecules with non-isolated H atoms \citep{ct85}.  
From laboratory studies of amorphous carbon particles, \citet{bb87} suggested that the 11.3\,$\mu$m feature is due to combined contributions of both OOP aromatic CH bending and infrared lattice modes.  
Through a spatial and spectral study of the Orion bar, \citet{ba89} concluded that
the narrow 11.3\,$\mu$m feature and the 11$-$13\,$\mu$m plateau emission 
are produced by separate components: free-flying PAHs and amorphous 
carbon particles or PAH clusters. 
Highly-hydrogenated PAH chrysene has also been proposed as the possible cause of the 11--13\,$\mu$m broad emission complex in F- and G-type stars  \citep{jb96}.  From a study of the broad emission plateau features observed in proto-planetary nebulae,  \citet{kv01} suggest that the 8 and 12-$\mu$m plateau features are due to superpositions of in-plane and OOP bending modes of aliphatic side groups attached to aromatic rings.  Quantum-chemical calculations indicate that the addition of aliphatic side groups indeed is able to produce plateau around 12\,$\mu$m \citep{sadjadi2015}.

In the next section, we will perform theoretical quantum chemistry calculations to determine the OOP bending mode frequencies of PAH molecules in the 11--12 $\mu$m region and explore the relationship between these modes and the astronomical 11.3 $\mu$m feature.

\section{Out-of-Plane C$-$H bending modes of PAH molecules}

\begin{figure}
\plotone{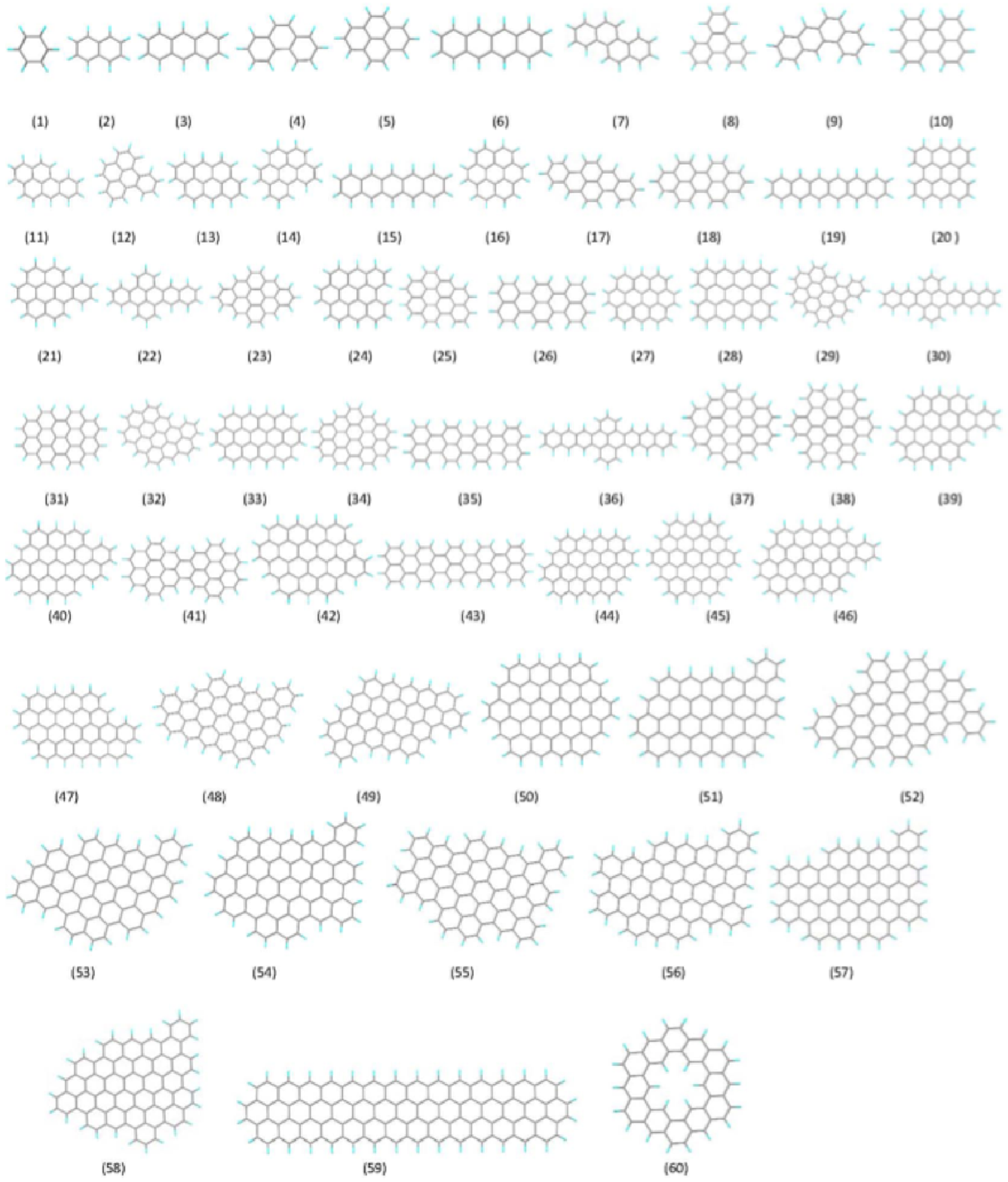}
\caption{Chemical structures of the 60 PAH molecules in our sample.  
The bracketed numbers below each structure correspond to the numbers in the first column of Table 1. 
\label{pah}}
\end{figure}

As the result of advances in quantum chemistry techniques, the vibrational spectrum of simple molecules such as PAH can be calculated very accurately.  In order to explore the OOP bending modes of PAH molecules, we have performed quantum chemistry calculations on a large group of PAH molecules.  A set of 60 neutral PAH molecules are selected from the PAH on-line databases of  \citet{boersma2014} and \citet{ malloci2007}.  The names and chemical formulae of these molecules are listed in Table 1 and their chemical strutures shown in Figure \ref{pah}.
We first use the  B3LYP  \citep{becke1993a, hertwig} and BHandHLYP hybrid functionals \citep{becke1993b}  in combination with polarization consistent basis set PC1  \citep{jensen2001, jensen2002}  to obtain 
the equilibrium geometries and the fundamental vibrational harmonic frequencies of the molecules. 
The calculations were based on density functional theory (DFT) using the Gaussian 09, Revision C.01 software package \citep{fri09} running on the  HKU grid-point supercomputer facility.  The B3LYP calculations were done using PQS \footnote{PQS version 4.0,  Parallel Quantum Solutions,   2013 Green Acres Road,  Fayetteville,  Arkansas  72703   URL: http://www.pqs-chem.com  Email:sales@pqs-chem.com: Parallel Quantum Solutions.} running on QS128-2300C-OA16 QuantumCubeTM machine.
The double scaling factors scheme of \citep{laury} were then applied to the DFT vibrational frequencies. In this scheme the vibrational frequencies $>1000$ cm$^{-1}$ and $<1000$ cm$^{-1}$ are scaled by 0.9311 and 0.9352 for BHandHLYP hybrid functionals  and 0.9654 and 0.9808 for B3LYP, respectively. The astronomical infrared  emission spectra are then simulated by applying the Drude model at $T$=500 K to these scaled vibrational lines.  The good accuracy and reliability of such DFT/Drude modeling in simulation of astronomical IR emission bands was demonstrated in  our previous study  \citep{sadjadi2015}.

All geometries have been optimized under the default criteria of the  cited ab intio quantum chemistry packages. The optimized geometries are all characterized as local minima, established by the positive values of all frequencies and their associated eigenvalues of the second derivative matrix. 
Visualizing and manipulating the results of vibrational normal mode analysis were performed by utilizing the Chemcraft program. Drude model is applied followed by a decomposition of normal mode analysis  \citep{sadjadi2015}.

\subsection{Dependence of the OOP frequencies on the number of exposed H sites}

\begin{figure}
\plotone{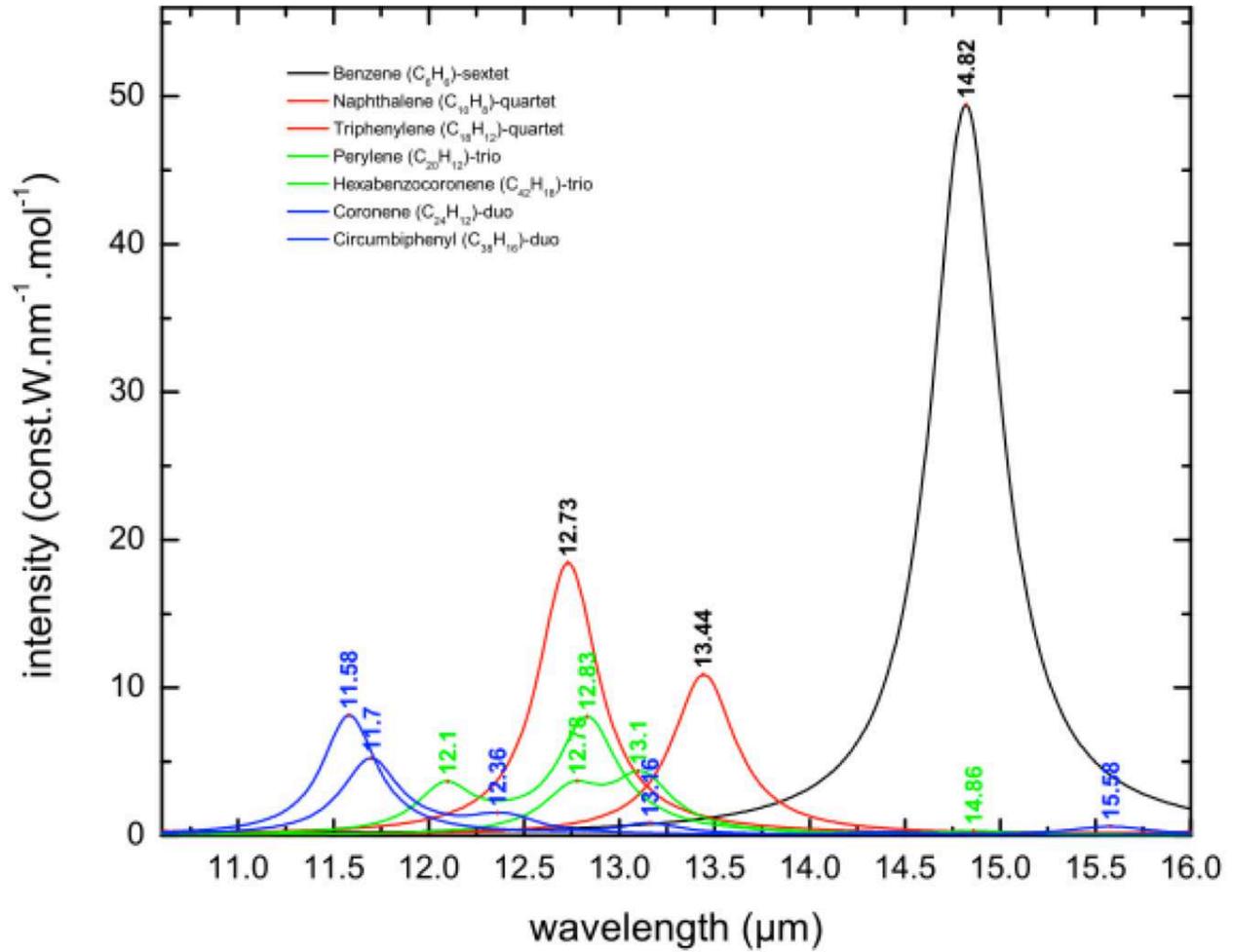}
\caption{Wavelengths of pure OOP bending modes of 7 simple PAH molecules with uniform edges.  Molecules with the same number of exposed edges are plotted in the same color (duo in red, trio in green, quartet in blue).  The benzene molecule (pure sextet, in black) is also plotted for comparison.  The peak wavelengths of each band are also labeled on the curves.
\label{solo}}
\end{figure}

In order to study the changes in frequency of  the OOP bending modes of molecules with different number of exposed H sites, we have selected 7 compact PAH molecules which have uniform edges, i.e. each of the molecule can be labeled as having only solo, duo, trio or quartet sites. Because of this uniformity, these molecules are expected to have less complicated  vibrational modes in the range of 10.8 to 15 $\mu$m (the accepted region of spectra for OOP vibrations).  Figure \ref{solo} shows the wavelengths of the pure OOP bending modes of these 7 molecules.  We can see that while the duo molecules (coronene and circumbiphenyl) have shorter wavelength peaks than the trio (perylene and hexabenzocoronene) and quartet (naphthalene and triphenylene) molecules, the latter two groups are less well separated.
While both naphthalene and triphenylene are classified as quartet sites, they show the OOP bands at 12.73 and 13.44 $\mu$m respectively. Perylene, which is classified as trio sites, shows the band at 12.83 $\mu$m close to that of naphthalene. 

The reason for this inconsistency can be explored by looking at the triphenylene (D3d) and perylene (D2h) optimized geometry and molecular symmetry. The Hs on each rings of these molecules are not equivalent as it is implied under the solo, duo, trio, quartet classification scheme. The Hs on triphenylene rings are grouped into two symmetry equivalent groups, i.e., 1 pair of duo H and two para solo Hs. In the same line of reasoning, the Hs on each of the rings on perylene are grouped into three different H atoms which are not symmetry equivalent of each other, although they are classified as trio.



For non-compact PAH molecules, there will be mixed solo, duo, trio and quartet sites.  Our quantitative normal mode displacement vector analysis yields the number of H atoms with major contributions in specific OOP vibrational mode. From these group of H atoms the number of solo, duo, trio and quartet H atoms are counted and converted into percentage values for the contribution of each edge-class in that OOP mode.  The percentage of solo, duo, trio and quartet sites of each of the 60 PAH molecules in our sample (including both uniform and non-uniform edge species) is given in Table 1.

\begin{figure}
\includegraphics[width=13cm]{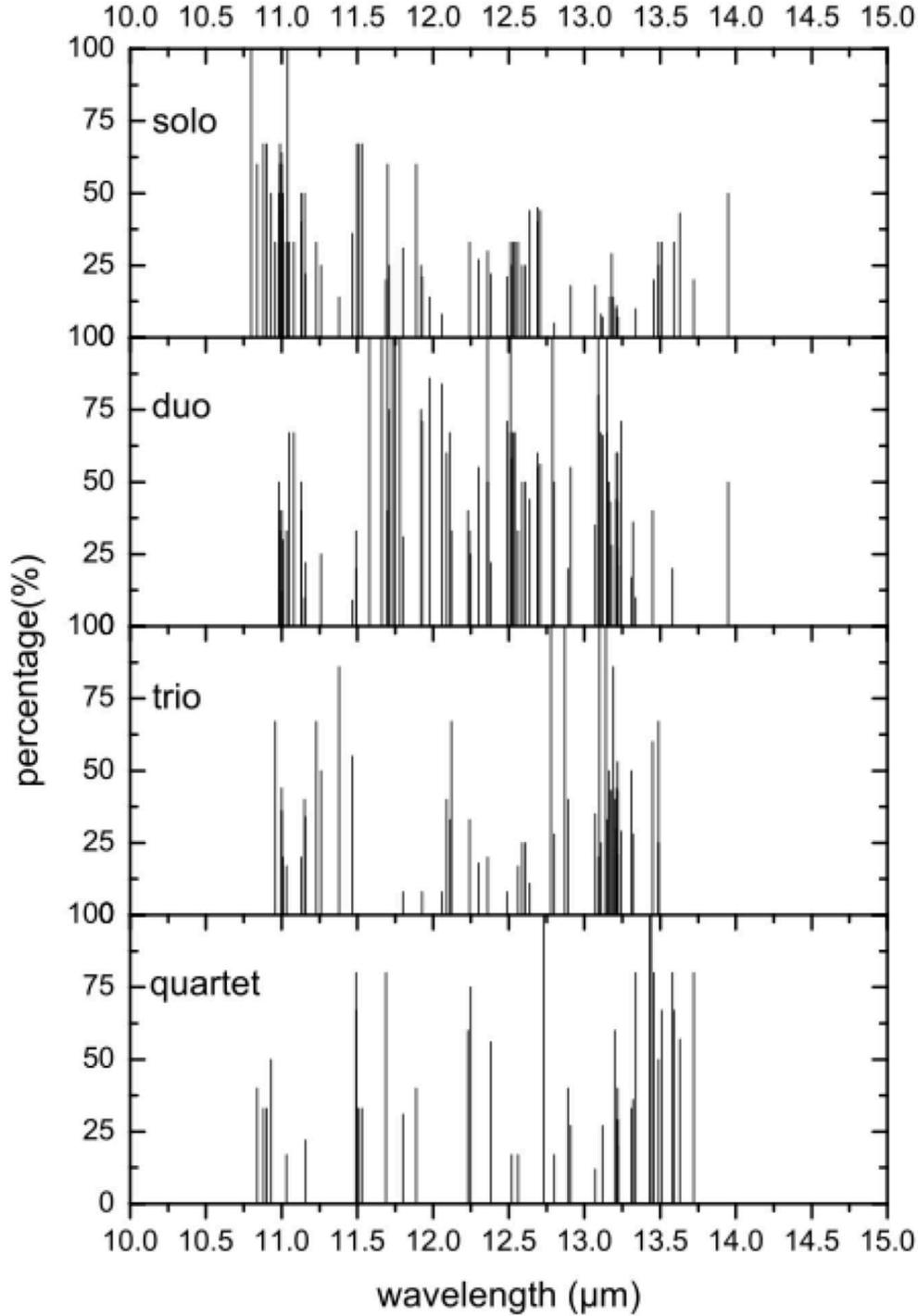}
\caption{The band positions of the 227 pure OOP modes of the 60 PAH molecules separated into peripheral H classes (from top to bottom: solo, duo, trio, and quartet).  The height of each line corresponds to the percentage of each class.}
\label{mixed}
\end{figure}

Out of the 227 vibrational bands in the 10.8--15 $\mu$m region, there are 112 bands that are found to be pure OOP vibrational modes, the other 115 are C--H OOP bands are coupled ring-OOP vibrations. Excluding the coupled OOP bands our vibrational analysis shows that all classes of H atoms could have significant contribution to the pure OOP bands, covering the entire wavelength range of 10.8 to 15 $\mu$m.   Figure \ref{mixed} plots the wavelength positions of each of the 112 pure OOP bands, separated into solo, duo, trio, and quartet classes (shown in four separate panels).  From the heights and concentrations of the lines, we can see that the solo bands are mostly clustered around 11.0 $\mu$m, although they can extend as far as 14 $\mu$m.  The duo modes are more evenly spread, with concentrations in particular around 11.8 $\mu$m, although there are also bands around 12.5 and 13.4 $\mu$m.  The trio bands are mostly concentrated around 13.2 $\mu$m.  The quattro modes are mostly concentrated around 13.5 $\mu$m but can be seen as short as 11 $\mu$m.


These results does not rule out the apparent connection between the OOPs frequencies and the number and kind of peripheral C-H bonds.  They show that correlation is not simple among different kinds of PAH molecule with different exposed edge. It is also even not simple if the number of exposed edge of a one PAH decrease by substituent.

In our previous study  \citep{sadjadi2015} on numbers of substituted ovalene structures, we have shown that the two OOP bands of core PAH at 11.11 and 11.78 $\mu$m do show the shifts when the number of edge C-H group decreases. However the bands are shifted in different ways. The 11.11 $\mu$m blue shifted to 10.98 $\mu$m not just because of the change in the number of exposed edges on the ring but because of the coupling to vibartional modes of aliphatic groups.
At the same time, the other strong OOP band at 11.78 $\mu$m is redshifted instead of blue shifted when the number of exposed edge is decreased. 

\subsection{PAH molecules as the carrier of the 11.3 $\mu$m feature}

\begin{figure}
\plotone{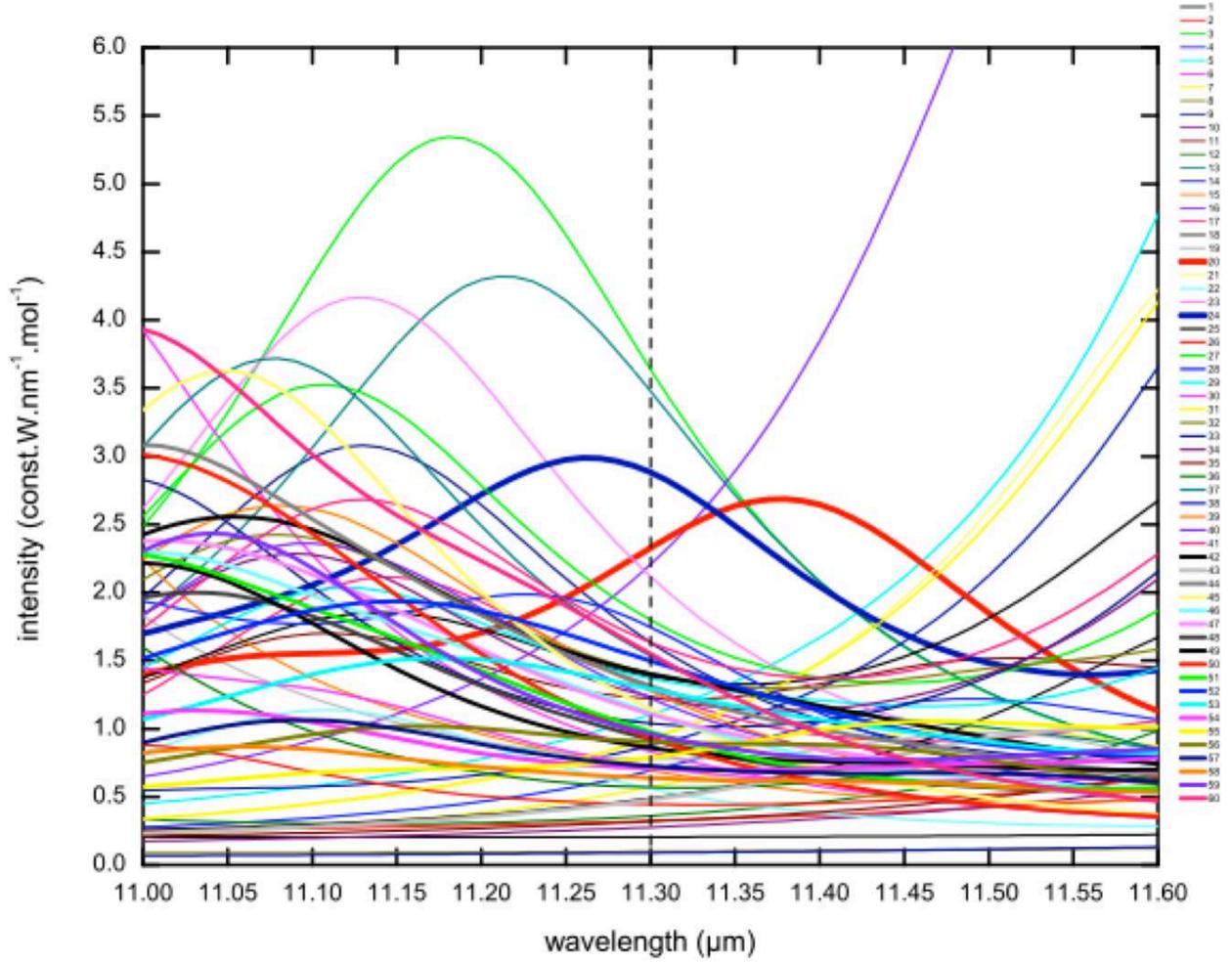}
\caption{Positions and profiles of the OOP bending modes of 60 PAH molecules.  The vertical dash line represents the nominal peak position of the 11.3 $\mu$m feature.
\label{graphene}}
\end{figure}

Since the OOP vibrational frequencies of PAH molecules can be calculated precisely, how well do they fit the astronomical 11.3 $\mu$m feature?  Our sample in Table 1 contains 60 neutral PAH molecules containing 6 to 80 C atoms of different honeycomb structures.  We have calculated the OOP bending mode frequencies of the PAH molecules. 
The simulated emission spectra of these molecules in the wavelength range  of 11 to 11.6 $\mu$m are plotted in Figure \ref{graphene}. We note that no band peaks are observed at the position of 11.30 $\mu$m among this set of PAH molecules.  Only two molecules in the sample, bisanthene (C$_{28}$H$_{14}$) and Dibenzo[bc,ef]coronene  (C$_{30}$H$_{14}$), exhibit a band close to 11.3 $\mu$m, at 11.38 and 11.26 $\mu$m, respectively.

Although 60 is not a large number, our results suggest that the  OOP bending modes of neutral PAH molecules, even including PAH molecules of different geometry with solo, duo, trio, etc. sites, are unlikely to be able  explain the observed astronomical 11.3 $\mu$m feature.  In the literature, support of the PAH hypothesis has cited the success in fitting the astronomical spectra using a mixture of PAH molecules with different sizes, geometry and charged states.  In the next section, we will explore in detail how well such fittings perform in fitting actual astronomical spectra.

\section{Fitting of the 11.3 $\mu$m  feature by the PAH model}

We performed spectral fitting of the observed 11.3\,$\mu$m feature in five astronomical objects using   the PAHdb and the IDL package {\it AmesPAHdbIDLSuite} developed by \citet{boersma2014} \footnote{http://www.astrochem.org/pahdb}. Two of the objects are planetary nebulae, one is a reflection nebula, one is a post-AGB star, and one is an active galaxy.   The astronomical spectra were downloaded from the {\it Spitzer} Heritage Archive and the {\it Infrared Space Observatory} ({\it ISO})  Archive. We have subtracted the continuum using a linear fit to the line-free regions.  Details of the fitting procedure can be found in \citet[][see their Sec.3]{bb13}.  Note that \citet{bb13} include only pure and nitrogen-containing PAHs.  In this study, all theoretical spectra contained in Version 2.00 of the PAHdb, including pure PAHs and heteroatom-molecules, was taken into account in the fitting. For the spectral fittings, we have employed the thermal approximation for all the species, in which the theoretical intensities were scaled by a Planck function at 500\,K. A Lorentzian profile with a band width of 15\,cm$^{-1}$ was assumed.

\begin{figure}
\plotone{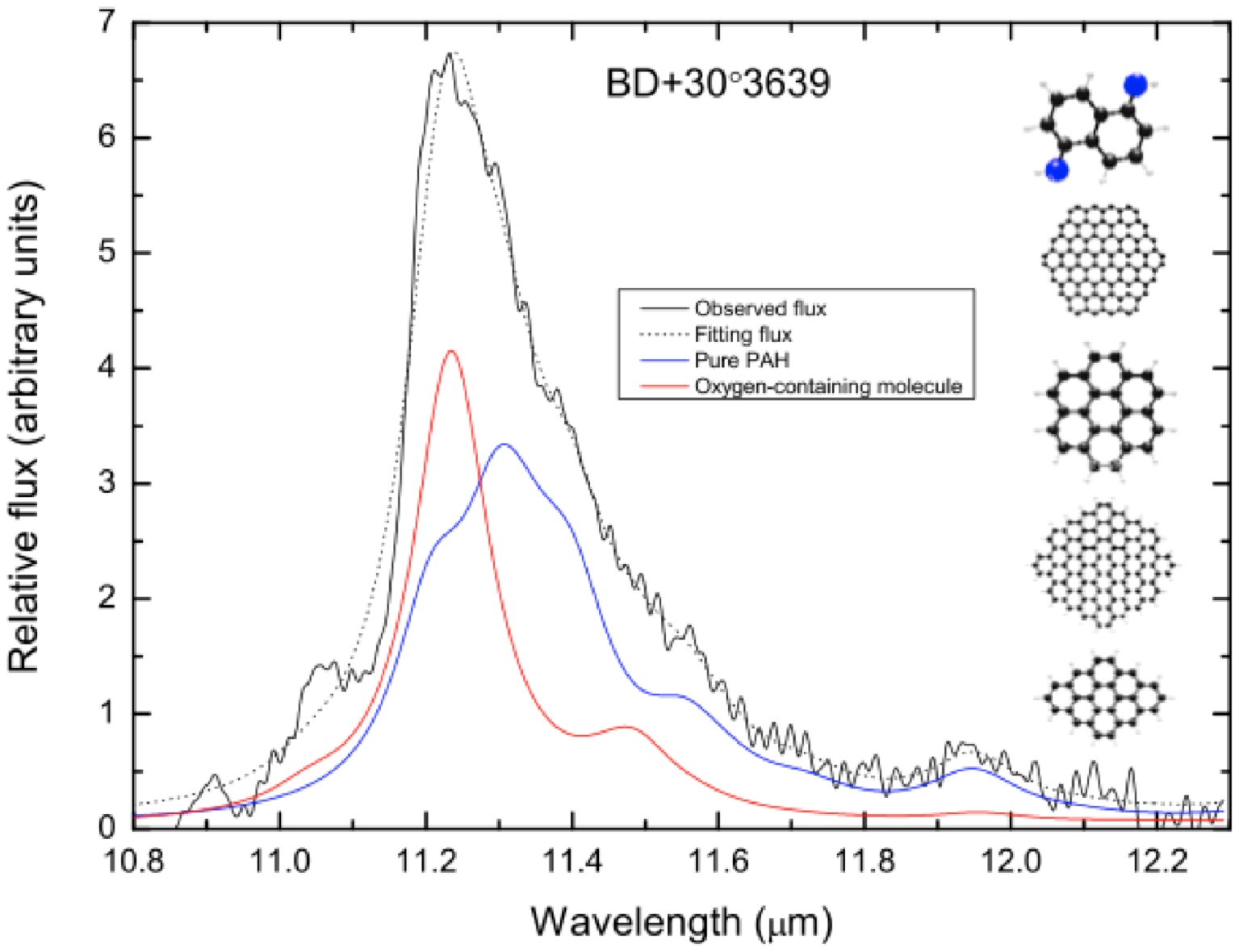}
\caption{Fittings of the 11.3 $\mu$m UIE feature in the planetary nebula BD+30\degr3639 using the PAHdb model.  The black solid line is the continuum-subtracted ISO spectrum and the black dotted line is the model fit by the PAHdb.  The blue line represents the total contributions from all the PAH ions and molecules, and the red line represents the contribution from oxygen-containing species.  The chemical structures of the major contributors are given on the right side of the figure (from top to bottom: C$_{10}$H$_{8}$O$_2$, C$_{96}^{+2}$, C$_{24}$H$_{12}^{+2}$, C$_{102}$H$_{26}$, and C$_{30}$H$_{14}$).  The C atoms are shown as black circles, H as small grey dots, and O atoms as blue circles.
\label{bd+30}}
\end{figure}

\begin{figure}
\plotone{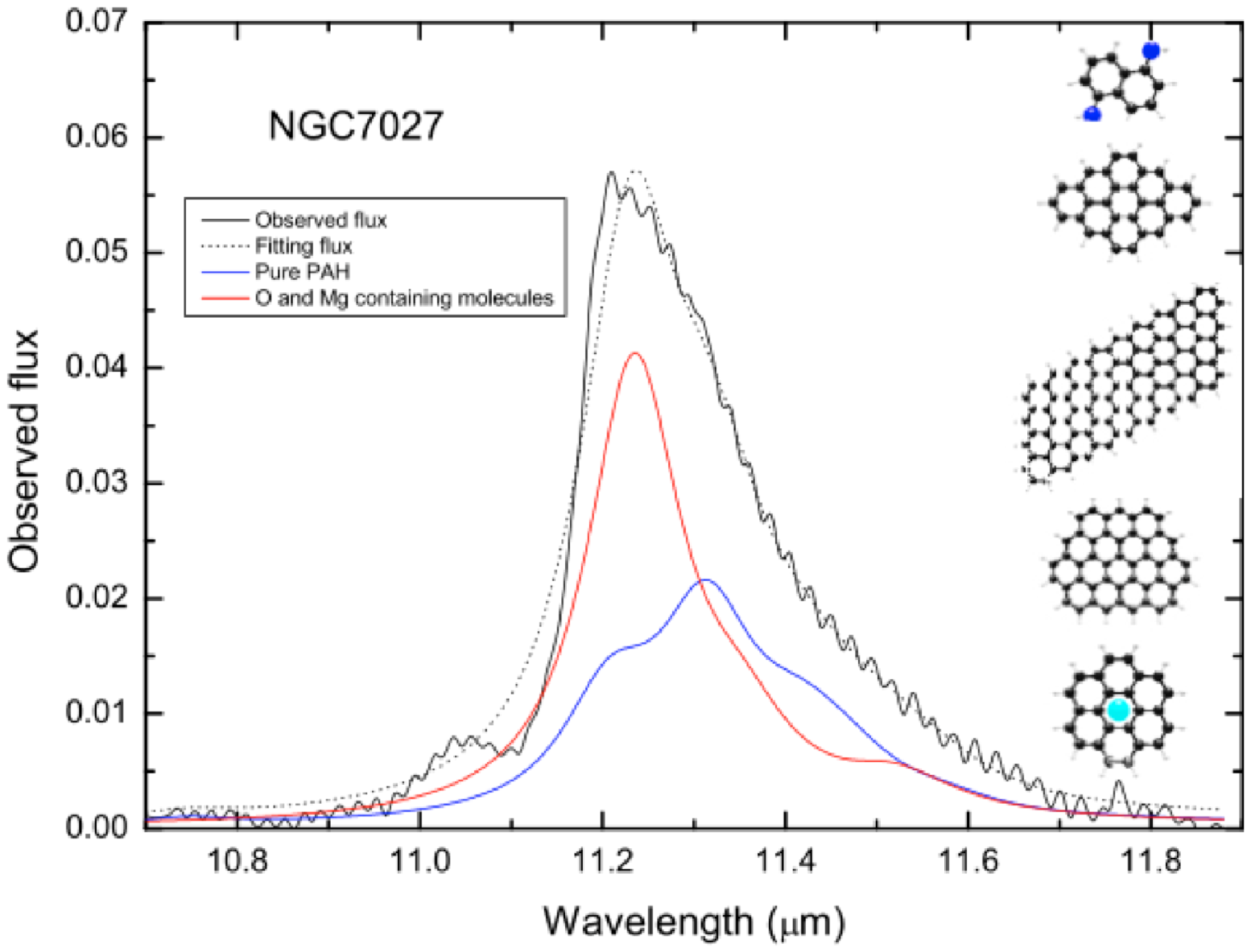}
\caption{Fittings of the 11.3 $\mu$m UIE feature in the planetary nebula NGC 7027 using the PAHdb model.  The black solid line is the continuum-subtracted ISO spectrum and the black dotted line is the model fit by the PAHdb.  The blue line represents the total contributions from all the PAH ions and molecules, and the red line represents the contribution from oxygen- and magnesium-containing species.  The chemical structures of the major contributors are given on the right side of the figure (from top to bottom: C$_{10}$H$_8$O$_2$, C$_{48}$H$_{18}$, C$_{98}$H$_{28}^+$, C$_{47}$H$_{17}$, C$_{24}$H$_{12}$Mg$^{+2}$).  The C atoms are shown as black circles, H as small grey dots, and O and Mg atoms as dark and light blue circles, respectively.
\label{ngc7027}}
\end{figure}

\begin{figure}
\plotone{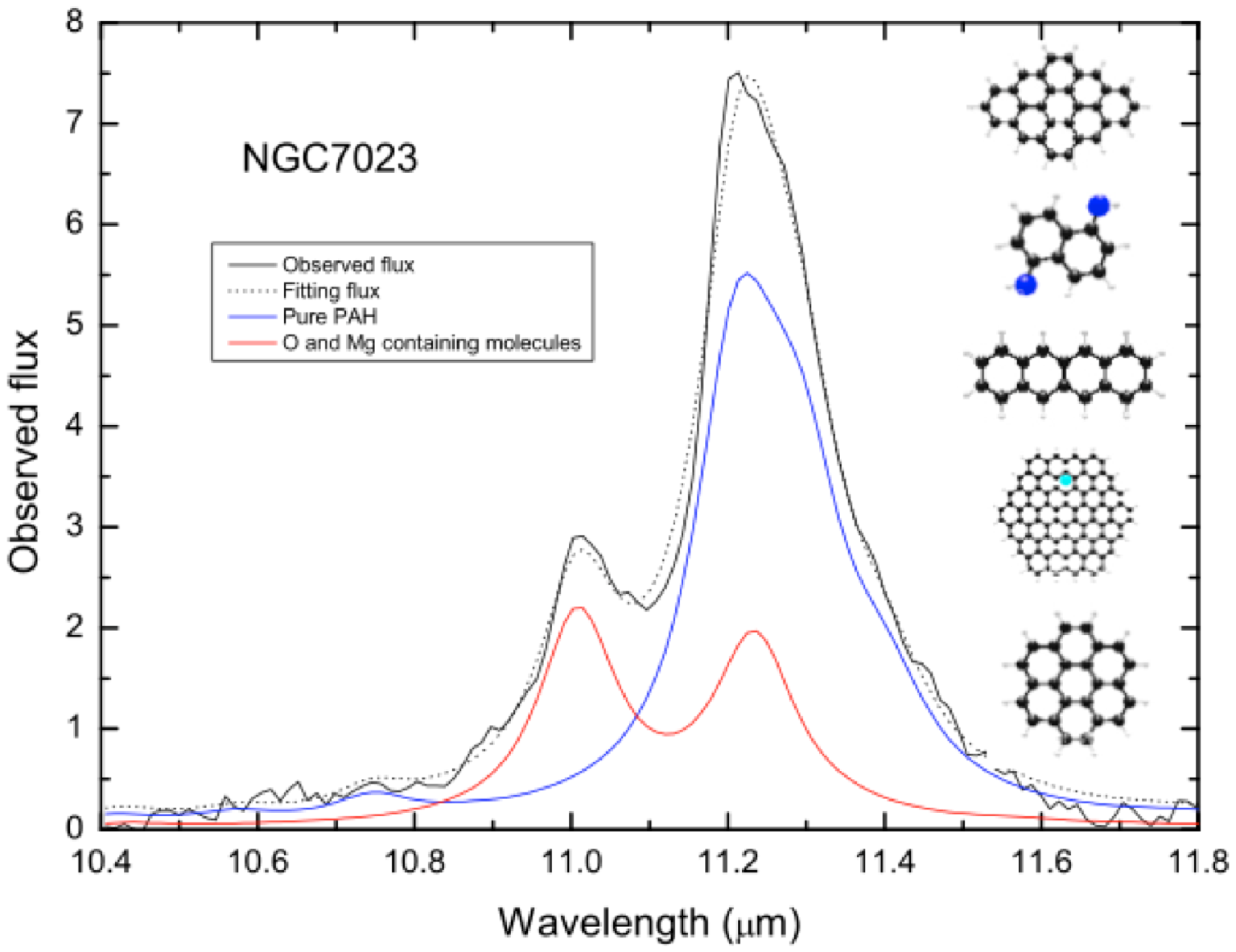}
\caption{Fittings of the 11.3 $\mu$m UIE feature in the reflection nebula NGC 7023 using the PAHdb model.  The black solid line is the continuum-subtracted Spitzer IRS spectrum and the black dotted line is the model fit by the PAHdb.    The blue line represents the total contributions from all the PAH ions and molecules, and the red line represents the contribution from oxygen- and magnesium-containing species.  The chemical structures of the major contributors are given on the right side of the figure (from top to bottom: C$_{48}$H$_{18}$, C$_{10}$H$_8$O$_2$, C$_{18}$H$_{12}$, C$_{96}$H$_{24}$Mg$^+$,  C$_{32}$H$_{14}$$^{+2}$).  The C atoms are shown as black circles, H as small grey dots, and O and Mg atoms as dark and light blue circles, respectively.
\label{ngc7023}}
\end{figure}

\begin{figure}
\plotone{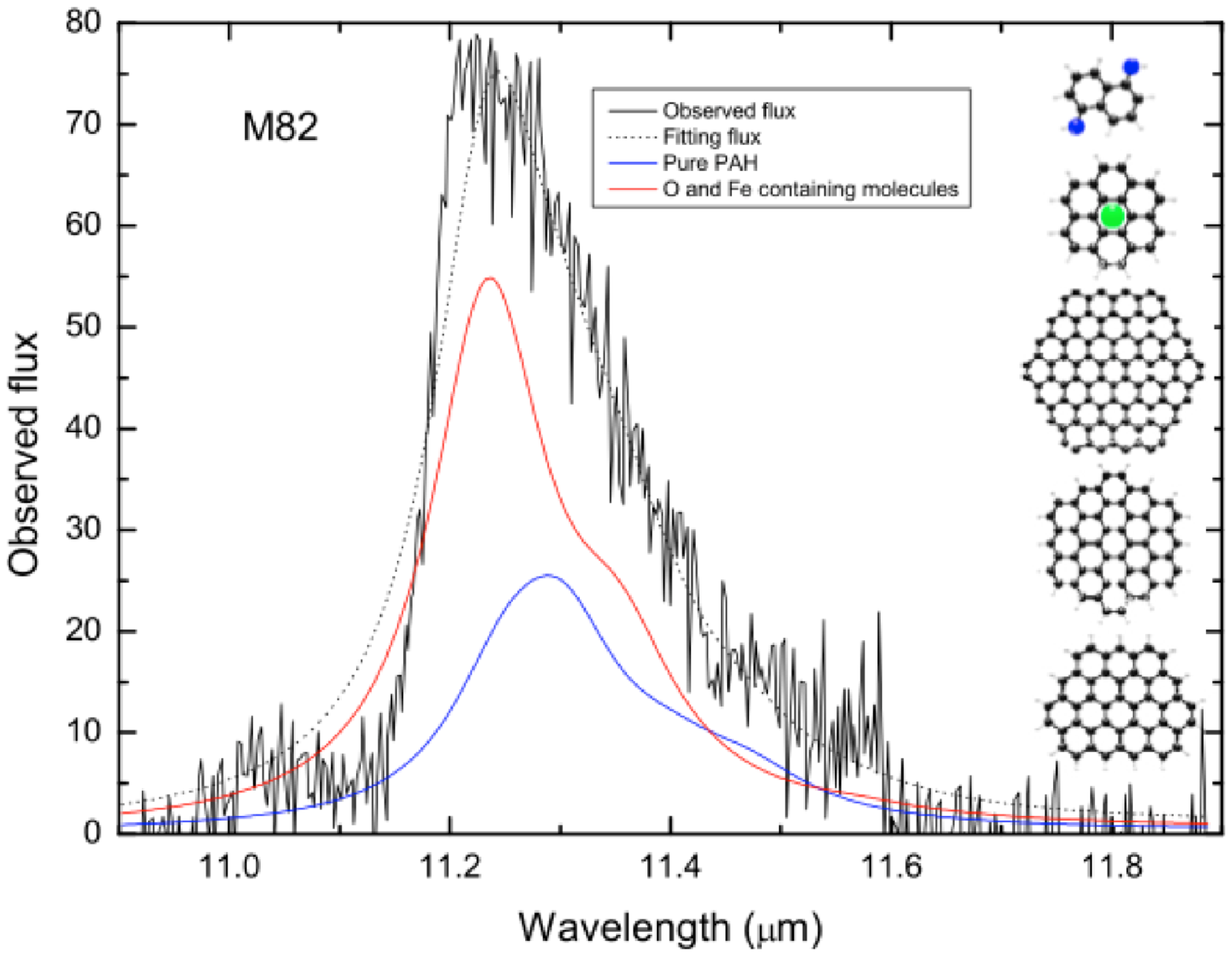}
\caption{Fittings of the 11.3 $\mu$m UIE feature in the active galaxy M82 using the PAHdb model.  The black solid line is the continuum-subtracted ISO spectrum and the black dotted line is the model fit by the PAHdb.   The blue line represents the total contributions from all the PAH ions and molecules, and the red line represents the contribution from oxygen- and magnesiaum-containing species.  The chemical structures of the major contributors are given on the right side of the figure (from top to bottom: C$_{10}$H$_8$O$_2$, C$_{24}$H$_{12}$Fe$^{+2}$, C$_{96}^{+2}$, C$_{54}$H$_{18}$, C$_{47}$H$_{17}$).  The C atoms are shown as black circles, H as small grey dots, O atoms as dark and Fe atoms as green circles.
\label{m82}}
\end{figure}

\begin{figure}
\plotone{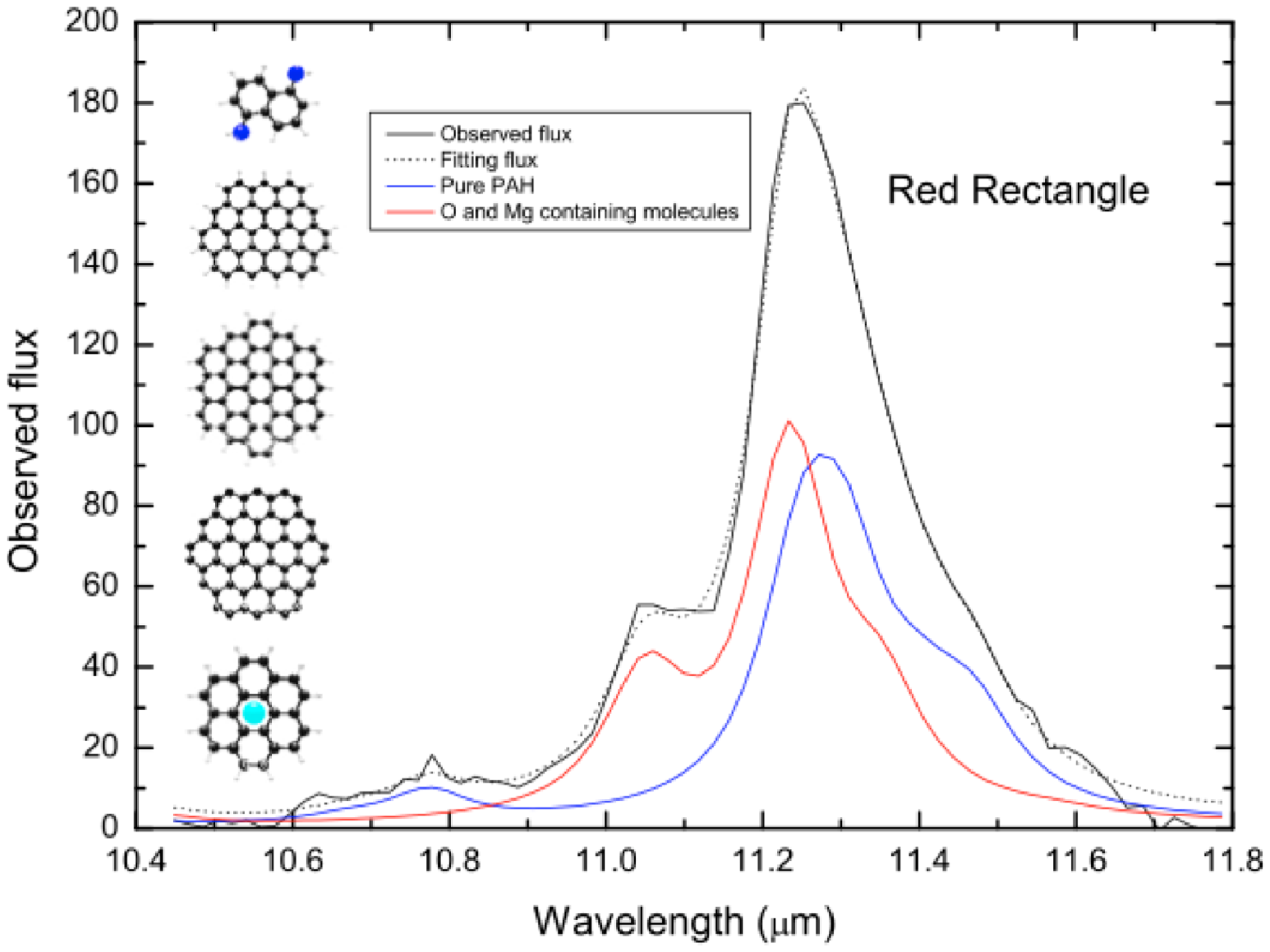}
\caption{Fittings of the 11.3 $\mu$m UIE feature in the reflection nebula HD 44179  using the PAHdb model.  The black solid line is the continuum-subtracted ISO spectrum and the black dotted line is the model fit by the PAHdb.   The blue line represents the total contributions from all the PAH ions and molecules, and the red line represents the contribution from oxygen- and magnesiaum-containing species.  The chemical structures of the major contributors are given on the right side of the figure (from top to bottom: C$_{10}$H$_8$O$_2$, C$_{47}$H$_{17}$, C$_{54}$H$_{18}$, C$_{54}$, C$_{24}$H$_{12}$Mg$^{+2}$).  The C atoms are shown as black circles, H as small grey dots, and O and Mg atoms as dark and light blue circles, respectively.
\label{rr}}
\end{figure}

     The fitting results are shown in Figures \ref{bd+30}--\ref{rr}.  The 5 molecular species that contribute the most to the fitting of the feature are displayed in each of the figures.  In all these examples, O-containing species (in particular dihydroynaphthalene C$_{10}$H$_8$O$_2$) is a major contributor to the 11.3 $\mu$m feature.   In some cases, its contribution is even larger than the total contribution from pure PAHs.  
Given the fact that the PAHdb contains hundreds of PAH molecules, it is surprising that O-containing (and sometimes Mg- and Fe-containing) molecules are needed to fit the 11.3 $\mu$m feature.

From our fittings, O-containing molecules contribute $42\%$,
$44\%$, $14\%$, $54\%$, and $45\%$ of the total flux of the
11.3\,$\mu$m feature in BD+30$^\circ$3639,  NGC\,7027, NGC\,7023, M\,82,
and HD\,44179, respectively. 
In principle, the fitting results can be used to estimate the abundance ratio of carbon locked in the O-containing molecules and pure PAHs if their UV absorption cross-sections are known. The calculation method is detailed in \citet{tielens2005} and has been applied to derive the abundances of C$_{60}$ and PAHs \citep{bt12}. However, to our knowledge, there is no reported measurement on the UV absorption cross-section of C$_{10}$H$_8$O$_2$. Adopting the same value as for PAH molecules ($7\times 10^{-18}$ cm$^{2}$ per C atom), we  assume that the fraction of carbon locked in C$_{10}$H$_8$O$_2$ and PAHs ($f$) is approximately equal to the ratio of their contributions to the flux of the 11.3\,$\mu$m feature. Using the median value of the flux contribution from O-containing species ($44\%$), we have a rough estimate of $f=0.8$, suggesting an O/C ratio of about 0.1.

We note that the Mg$^{2+}$ and Fe$^{2+}$ containing molecules are similar complexes where the PAH is acting as $\pi$ electron donor to the metal atoms.  Also, the molecules are selected by PAHdb to achieve a good fit and they are not necessarily physically viable molecules.  For example, C$_{54}$ (used to fit the Red Rectangle) and C$_{96}^+$ (used to fit M82) have very high number of dangling bonds and should be extremely reactive.  It is unrealistic to expect molecules of very short lifetimes to play a role in the formation of the astronomical 11.3 $\mu$m feature in the interstellar medium.

Since the code automatically selects spectra from the data pool to get the best fitting and the data pool contains many hundreds of  pure PAHs and very few heteroatom molecules, we can safely conclude
that the inclusion of the O-containing species is needed to achieve a good fit.  We can compare these fits with the fits to NGC 7023 by \citet{bb13}.  Examination of Figure 4 of \citet{bb13} shows that their  model spectrum contains features on the shoulders of the 11.3 $\mu$m feature, which are not seen in observations.  This confirms that the PAHdb model, even with the large number of parameters, has difficulty fitting the feature with pure PAH molecules and ions.

\section{Discussions}

A successful model for the carrier of  the UIE phenomenon has to produce strong consistent features at 3.3, 6.2, 7.7, 8.6, and 11.3 $\mu$m, accompanied by broad plateau features around 8 and 12 $\mu$m, and have no strong narrow features in between the strong UIE bands.  The difficulty in PAH molecules meeting these criteria has been discussed by \citet{cook1998}.  The way  to get around these problems in the past is to invoke a large mix of PAH molecules of different masses, sizes, geometry and ionized states and fit the astronomical spectra by adjusting the relative abundance of hundreds of molecular species.  Anharmonic shifts and hot bands are also included to improve the fits \citep{joblin2002}.  By introducing an emission model that includes a mass distribution of PAH molecules, temperature-dependent linewidths, and the spectral energy distribution of the background radiation source, \citet{candian2015} are able to fit the observed profiles of the 11.3 $\mu$m feature.  For example, the asymmetric profile of the feature is reproduced by having different mass components.   Recent work has expanded the PAH database to incorporate dehydrogenated PAH molecules,  including those with non-planar structures \citep{mackie2015}.   
Another possibility is to include protonated PAHs.  However, recent experimental IR spectra suggest that neither  neutral-protonated (H.PAH) or cationic-protonated  (H$^+$PAH) forms  could successfully produce the 11.3 $\mu$m feature in moderate-size PAHs such as coronene \citep{bahou}.
Since astronomical UIE spectra are observed in very different radiation environments, it would not be reasonable to assume that the PAH population will consist of PAH molecules of the same mix of physical (ionization, dehydrogenation, protonation) states and mass distributions everywhere. 
The large number of free parameters used in these fits also raise doubts about their validity as practically any spectra can be fitted by such methods \citep{zk15}.

\begin{figure}
\plotone{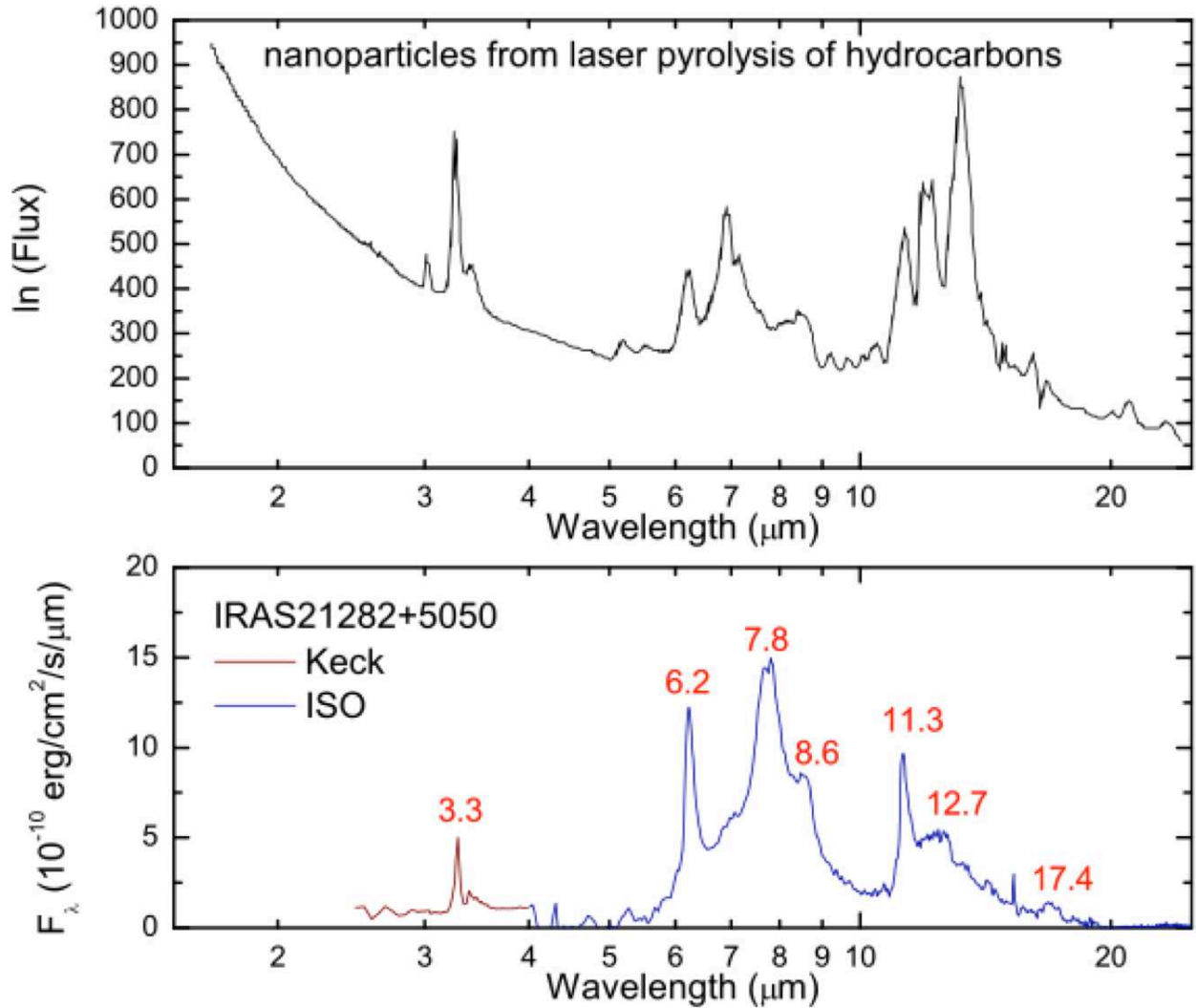}
\caption{Comparison of the  laboratory spectrum of nanoparticles produced by laser pyrolysis of hydrocarbons \citep{herlin1998} (top panel) with the astronomical spectrum of the planetary nebula IRAS 21282+5050 (bottom panel).  The peak wavelengths (in units of $\mu$m) of the astronomical UIE bands are marked in red.  The long wavelength ($>$4 $\mu$m) part of the spectrum of IRAS 21282+5050 is from {\it ISO} and the short wavelength ($<$4 $\mu$m) are from ground-based observations \citep{hrivnak2007}.}
\label{herlin}
\end{figure}
\begin{figure}
\plotone{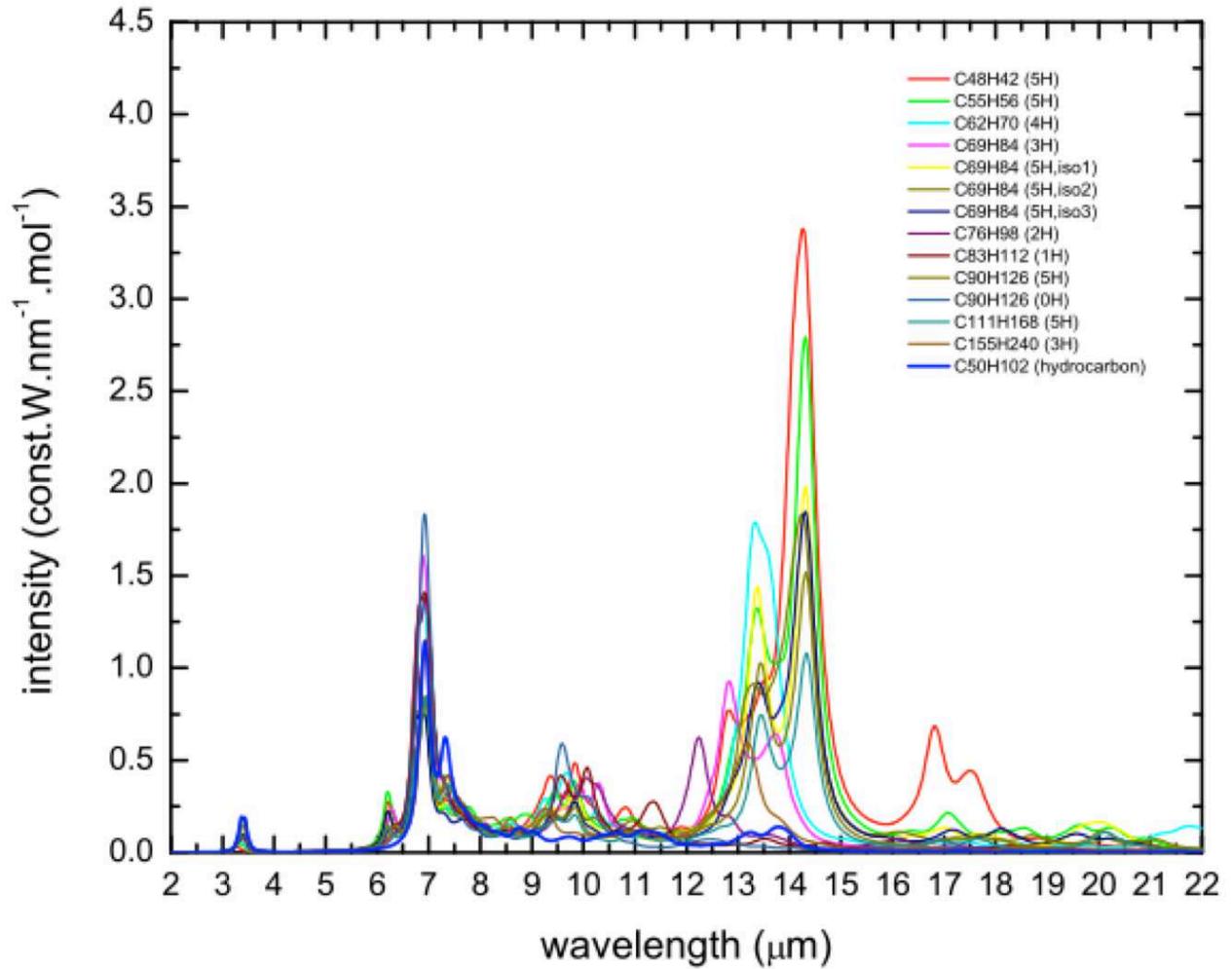}
\caption{The infrared spectra of 13 simple MAON-like molecules.  In spite of the different sizes of the molecules, the vibrational modes seem to cluster around consistent frequencies.  The spectrum of the pure aliphatic molecule C$_{50}$H$_{102}$ is also plotted for comparison.}  
\label{maon}
\end{figure}
\begin{figure}
\plotone{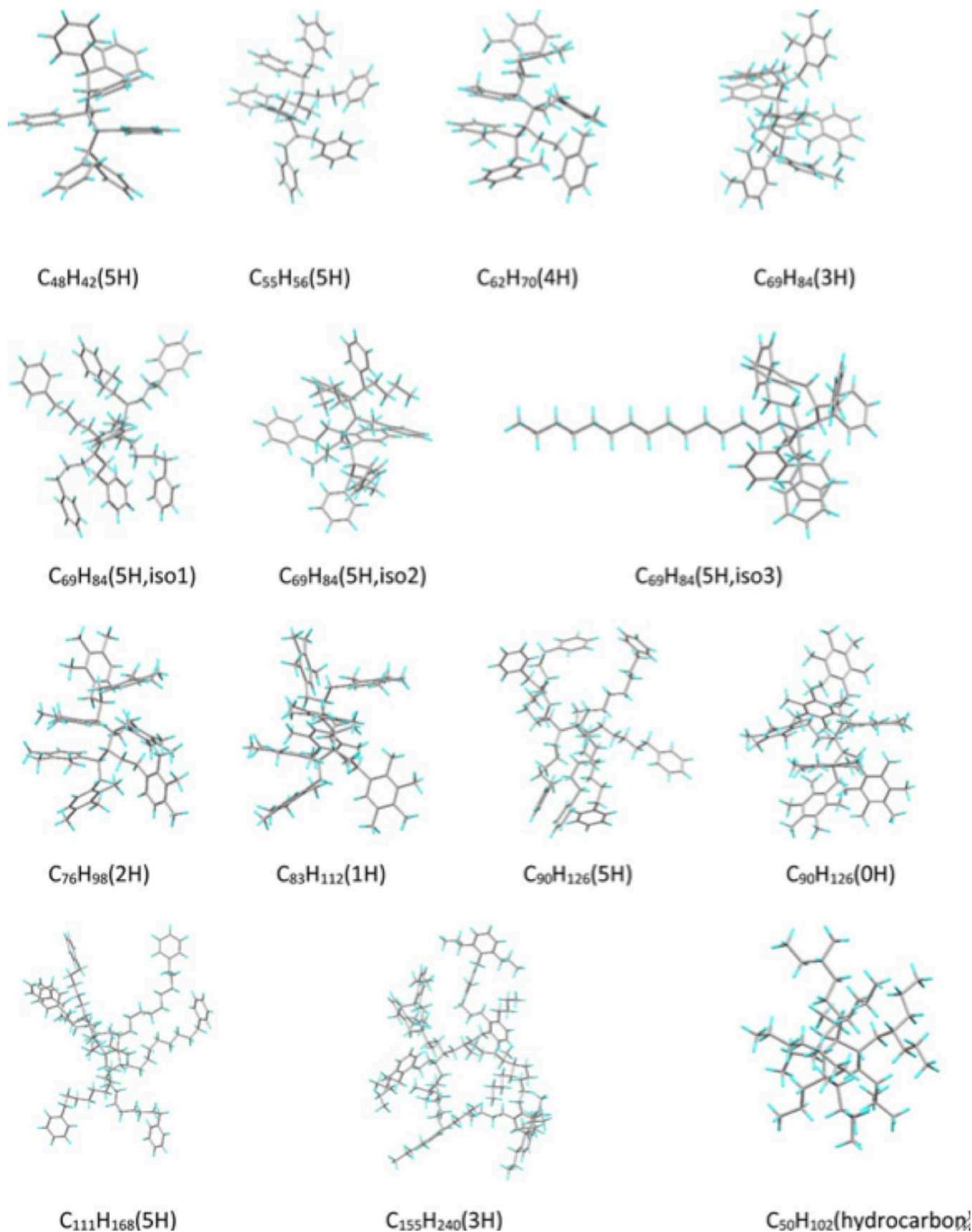}
\caption{Chemical structures of the 14 molecules used in Fig.~\ref{maon}.  All the aromatic rings in each of the molecule has the same number of H not substituted by methyl and methylene groups (the number inside the brackets).  One molecule (C$_{50}$H$_{102}$) has no aromatic rings.  The labels iso1, iso2, and iso3 refer to three different isomers of the molecule C$_{69}$H$_{84}$.}
\label{maonchem}
\end{figure}

If PAH molecules are not the carriers of the 11.3 $\mu$m feature, what are the other possible candidates?  
Amorphous carbonaceous solids are also known to possess infrared vibrational based similar to the UIE bands \citep{dischler1983a, dischler1983b, guillois1996}.
Figure \ref{herlin} shows a comparison between the astronomical UIE spectrum with the laboratory spectrum of nanoparticles produced by laser pyrolysis of hydrocarbons \citep{herlin1998}.   The laboratory spectrum of these particles show general resemblance to the astronomical spectrum, not only in the major UIE bands, but also in the plateau features.  The features are naturally broad and do not need to be artificially broadened as in gas-phase molecules.  In addition to the 11.3 $\mu$m feature, strong peaks are seen at 11.9 and 13.3 $\mu$m,


In order to examine the infrared activity of hydrocarbons with mixed aromatic/aliphatic structures, we have calculated the simulated infrared spectra (Fig.~\ref{maon}) for 13 different examples of such structures (Fig.~\ref{maonchem}).  For simplicity, we have only explored molecules that consist of single aromatic rings and aliphatic chains consisting of methyl and methylene groups.   As such, they represent a simple subset of MAONs \citep{kz11}.   The number of free hydrogens on all benzene rings (the numbers inside the bracket in Fig.~\ref{maon}) is kept the same in each structure. We also include an example of pure aliphatic molecule (C$_{50}$H$_{102}$) for comparison.

The first notable difference between these molecules and PAH molecules is in the way that the infrared bands appear at different parts of spectra as the size of molecules increases. The infrared bands of PAH molecules with different sizes can appear in every part of the spectra as each vibrational mode has significant variations in peak wavelengths.   The MAON-like spectra, however, have vibrational modes occurring at consistent wavelengths and are not highly dependent on the size of the molecules (Fig.~\ref{maon}). 
The 6 main features around 3.4, 6.2--6.4, 6.9--7.4, 9--12, 12--15, 15--20 $\mu$m can roughly be identified as aliphatic C--H stretching, aromatic C=C stretching, scissors mode of methylene C--H bonds coupled with umbrella deformation mode of methyl C-H bonds, methyl and methylene deformation modes coupled with aromatic C-H in plane bending modes, aromatic C--H OOPs, aliphatic fragment group frequencies coupled with aromatic ring OOP deformation modes. However as discussed by \citet{sadjadi2015}, all of these modes have varying degree of coupling between aromatic and aliphatic components.
The 6 peaks in Figure \ref{maon} show some qualitative resemblance to the astronomical UIE spectra (lower panel, Figure \ref{herlin}), although not exact correspondence in peak wavelengths.  
For example,  the OOP bending modes of these 14 molecules range from 12 to 15 $\mu$m approximately resemble the 12 $\mu$m emission plateau feature but do not show a definite peak at 11.3 $\mu$m.  Further theoretical investigation on the vibrational modes of MAON structures with larger sizes and complexity is needed.



\section{Conclusions}

The UIE bands are commonly found throughout the Universe and the energy emitted from these bands represent a significant output of active galaxies.  Since the strengths of UIE bands and their distributions are used as a diagnostic tool of the physical and radiative environment of galactic and extragalactic objects, a correct identification of its origin is of utmost importance.  Whether it originates from gas-phase molecules or complex organic solids can also alter our view of the chemical abundance of interstellar matter.  

The astronomical UIE bands have consistent patterns in different astronomical environments and such consistency is difficult to explain by simple, gas-phase PAH molecules.  The 11.3 $\mu$m UIE feature cannot be fitted by superpositions of hundreds of pure PAH molecules, even including PAH molecules of different sizes and  charged states.  In order to correctly identify the origin of the UIE bands, we need to explore more complex structures, specifically those with mixed aromatic/aliphatic structures.  Preliminary results in this paper show that the MAON-like structures have consistent clusters of vibrational bands and therefore are viable candidates for the carrier of the astronomical UIE bands.

{\flushleft \bf Acknowledgment~}

\acknowledgments
We dedicate this paper to our friend and colleague Tom Ziegler who passed away in 2015.  Tom was a pioneer in the development of the DTF method which allows the calculations performed in this paper.
We thank Franco Cataldo and Renaud Papoular for helpful comments on an earlier draft of this manuscript.  The fitting models were done using the NASA Ames Research Center PAH IR Spectroscopic Database and software and we thank Christiaan Boersma  for making these resources publicly available. This work was partially supported by the Research Grants Council of the Hong Kong Special Administrative Region, China (project no. 17302214).

\clearpage

\begin{deluxetable}{lcccccccc}
\tablecaption{Our sample of PAH molecules and the peripheral H sites in each molecule\label{sample}}
\tabletypesize{\scriptsize}
\tablehead{
\colhead{No.} & \colhead{Molecule} & \colhead{Formula} & \multicolumn{6}{c}{percentages}\\
\cline{4-9}
& & & \colhead{solo} & \colhead{duo}  & \colhead{trio} & \colhead{quartet}&\colhead{quintet}&\colhead{sextet}
}
\startdata
1  &Benzene &C$_{6}$H$_{6}$  &0 &0 &0 &0 &0 &100 \\ 
2  &Naphthalene & C$_{10}$H$_{8}$ & 0 & 0 & 0 & 100 & 0 & 0 \\ 
3  &Anthracene & C$_{14}$H$_{10}$ & 20 & 0 & 0 & 80 & 0 & 0 \\ 
4  &phenanthrene & C$_{14}$H$_{10}$ & 0 & 20 & 0 & 80 & 0 & 0 \\ 
5  &Pyrene & C$_{16}$H$_{10}$ & 0 & 40 & 60 & 0 & 0 & 0 \\ 
6  &Tertracene & C$_{18}$H$_{12}$ & 33 & 0 & 0 & 67 & 0 & 0 \\ 
7  &Chrysene & C$_{18}$H$_{12}$ & 0 & 33 & 0 & 67 & 0 & 0 \\
8  &Triphenylene & C$_{18}$H$_{12}$ & 0 & 0 & 0 & 100 & 0 & 0 \\ 
9  &Benzo[a]anthracene & C$_{18}$H$_{12}$ & 17 & 17 & 0 & 66 & 0 & 0 \\ 
10 &Perylene &C$_{20}$H$_{12}$ & 0 & 0 & 100 & 0 & 0 & 0 \\ 
11 &Benzo[a]pyrene &C$_{20}$H$_{12}$ & 9 & 33 & 25 & 33 & 0 & 0 \\
12 &Benzo[e]pyrene &C$_{20}$H$_{12}$&0 & 17 & 50 & 33 & 0 & 0 \\ 
13 &Anthanthrene &C$_{22}$H$_{12}$&17 & 33 & 50 & 0 & 0 & 0 \\ 
14 &Benzo[ghi]perylene &C$_{22}$H$_{12}$&0 & 50 & 50 & 0 & 0 & 0 \\
15 &Pentacene &C$_{22}$H$_{14}$&43 & 0 & 0 & 57 & 0 & 0 \\
16 &Coronene &C$_{24}$H$_{12}$&0 & 100 & 0 & 0 & 0 & 0 \\
17 &Dibenzo[b,def]chrysene &C$_{24}$H$_{14}$&14 & 29 & 0 & 57 & 0 & 0 \\ 
18 &Dibenzo[cd,lm]perylene &C$_{26}$H$_{14}$&0 & 57 & 43 & 0 & 0 & 0 \\ 
19 &Hexacene &C$_{26}$H$_{16}$&50 & 0 & 0 & 50 & 0 & 0 \\
20 &Bisanthene &C$_{28}$H$_{14}$&14 & 0 & 86 & 0 & 0 & 0 \\ 
21 &Benzo[a]coronene &C$_{28}$H$_{14}$&0 & 71 & 0 & 29 & 0 & 0 \\
22 &Dibenzo[fg,st]pentacene &C$_{28}$H$_{16}$&13 & 0 & 37 & 50 & 0 & 0 \\
23 &Dibenzo[bc,kl]coronene &C$_{30}$H$_{14}$&29 & 29 & 42 & 0 & 0 & 0 \\
24 &Dibenzo[bc,ef]coronene &C$_{30}$H$_{14}$&14 & 43 & 43 & 0 & 0 & 0 \\
25 &Naphtho[8,1,2abc]coronene &C$_{30}$H$_{14}$&7 & 71 & 22 & 0 & 0 & 0 \\
26 &Terrylene &C$_{30}$H$_{16}$&0 & 25 & 75 & 0 & 0 & 0 \\ 
27 &Ovalene &C$_{32}$H$_{14}$&14 & 86 & 0 & 0 & 0 & 0 \\
28 &Tetrabenzocoronene &C$_{36}$H$_{16}$&25 & 0 & 75 & 0 & 0 & 0 \\
29 &Benz[a]ovalene &C$_{36}$H$_{16}$&13 & 62 & 0 & 25 & 0 & 0 \\ 
30 &Dibenzo[hi,yz]heptacene  &C$_{36}$H$_{20}$&30 & 0 & 30 & 40 & 0 & 0 \\
31 &Circumbiphenyl &C$_{38}$H$_{16}$&0 & 100 & 0 & 0 & 0 & 0 \\ 
32 &Naphth[8,2,1,abc]ovalene &C$_{38}$H$_{16}$&19 & 62 & 19 & 0 & 0 & 0 \\ 
33 &Circumanthracene &C$_{40}$H$_{16}$&25 & 75 & 0 & 0 & 0 & 0 \\ 
34 &Phenanthro[3,4,5,6 vuabc]ovalene &C$_{40}$H$_{16}$&13 & 87 & 0 & 0 & 0 & 0 \\ 
35 &Quaterrylene &C$_{40}$H$_{20}$&0 & 40 & 60 & 0 & 0 & 0 \\ 
36 &Dibenz[jk,a1b1]octacene &C$_{40}$H$_{22}$&36 & 0 & 28 & 36 & 0 & 0 \\
37 &Circumpyrene &C$_{42}$H$_{16}$&25 & 75 & 0 & 0 & 0 & 0 \\
38 &Hexabenzocoronene &C$_{42}$H$_{18}$&0 & 0 & 100 & 0 & 0 & 0 \\
39 &Honeycomb15$^a$ &C$_{46}$H$_{18}$&22 & 56 & 0 & 22 & 0 & 0 \\
40 &Honeycomb16 &C$_{48}$H$_{18}$&17 & 66 & 17 & 0 & 0 & 0 \\ 
41 &Dicoronylene &C$_{48}$H$_{20}$&20 & 80 & 0 & 0 & 0 & 0 \\ 
42 &Honeycomb17 &C$_{50}$H$_{18}$&28 & 55 & 17 & 0 & 0 & 0 \\ 
43 &Pentarylene &C$_{50}$H$_{24}$&0 & 50 & 50 & 0 & 0 & 0 \\
44 &Honeycomb18 &C$_{52}$H$_{18}$&33 & 67 & 0 & 0 & 0 & 0 \\
45 &Circumcoronene &C$_{54}$H$_{18}$&33 & 67 & 0 & 0 & 0 & 0 \\
46 &Honeycomb19 &C$_{56}$H$_{20}$&30 & 50 & 0 & 20 & 0 & 0 \\ 
47 &Honeycomb20 &C$_{58}$H$_{20}$&35 & 50 & 15 & 0 & 0 & 0 \\
48 &Honeycomb21 &C$_{62}$H$_{22}$&32 & 36 & 14 & 18 & 0 & 0 \\
49 &Honeycomb22 &C$_{64}$H$_{22}$&36 & 36 & 28 & 0 & 0 & 0 \\
50 &Circumovalene &C$_{66}$H$_{20}$&40 & 60 & 0 & 0 & 0 & 0 \\
51 &Honeycomb23 &C$_{66}$H$_{22}$&36 & 46 & 0 & 18 & 0 & 0 \\ 
52 &Honeycomb24 &C$_{70}$H$_{24}$&25 & 33 & 25 & 17 & 0 & 0 \\
53 &Honeycomb25 &C$_{72}$H$_{24}$&21 & 50 & 12 & 17 & 0 & 0 \\
54 &Honeycomb26 &C$_{74}$H$_{24}$&21 & 50 & 12 & 17 & 0 & 0 \\ 
55 &Honeycomb27 &C$_{78}$H$_{26}$&19 & 31 & 35 & 15 & 0 & 0 \\
56 &Honeycomb28 &C$_{80}$H$_{26}$&19 & 31 & 35 & 15 & 0 & 0 \\
57 &Honeycomb29 &C$_{82}$H$_{26}$&19 & 31 & 35 & 15 & 0 & 0 \\
58 &Honeycomb30 &C$_{84}$H$_{26}$&15 & 47 & 23 & 15 & 0 & 0 \\ 
59 &Honeycomb43 &C$_{120}$H$_{36}$&67 & 33 & 0 & 0 & 0 & 0 \\
60 &Kekulene    &C$_{48}$H$_{24}$&50 & 50 & 0 & 0 & 0 & 0 \\  
\enddata
\tablenotetext{a}{The number "xx" in the label "Honeycombxx" denotes the number of fused benzene rings in the molecular structure of the PAH molecule.}
\end{deluxetable}

\end{CJK*}

\end{document}